\RequirePackage{fix-cm}
\documentclass[natbib,smallextended]{svjour3}       

\smartqed  
\usepackage{graphicx}
\usepackage{amssymb}
\usepackage{amsmath}
\usepackage{mathabx}
\usepackage{xspace}
\usepackage{hyperref}
\newcommand{\Conv}{\mathop{\scalebox{1.5}{\raisebox{-0.2ex}{$\ast$}}}}%
\newcommand{\Euclid}{\textit{Euclid}\xspace}
\renewcommand{\deg}{\ensuremath{^{\circ}}\xspace}
 \journalname{Experimental Astronomy}

\begin{document}

\title{Measuring a Charge-Coupled Device Point Spread Function
}
\subtitle{\Euclid Visible Instrument CCD273-84 PSF Performance}

\author{Sami-Matias Niemi       \and
        Mark Cropper \and
        Magdalena Szafraniec \and
        Thomas Kitching
}

\institute{S.-M. Niemi \at
              Mullard Space Science Laboratory, University College London, Holmbury St Mary, Dorking, Surrey, RH5 6NT, UK \\
              \email{s.niemi@ucl.ac.uk} 
           \and
          M. Cropper \at
              Mullard Space Science Laboratory, University College London, Holmbury St Mary, Dorking, Surrey, RH5 6NT, UK \\
              \email{m.cropper@ucl.ac.uk} 
           \and
           M. Szafraniec \at
              Mullard Space Science Laboratory, University College London, Holmbury St Mary, Dorking, Surrey, RH5 6NT, UK \\
              \email{m.szafraniec@ucl.ac.uk}  
           \and
           T. Kitching \at
              Mullard Space Science Laboratory, University College London, Holmbury St Mary, Dorking, Surrey, RH5 6NT, UK \\
              \email{t.kitching@ucl.ac.uk} 
}

\date{Received: date / Accepted: date}

\maketitle

\begin{abstract}

In this paper we present the testing of a back-illuminated development \Euclid Visible Instrument (VIS) Charge-Coupled Device (CCD) to measure the intrinsic CCD Point Spread Function (PSF) characteristics using a novel modelling technique. We model the optical spot projection system and the CCD273-84 PSF jointly. We fit a model using Bayesian posterior probability density function, sampling to all available data simultaneously. The generative model fitting is shown, using simulated data, to allow good parameter estimations even when these data are not well sampled. Using available spot data we characterise a CCD273-84 PSF as a function of wavelength and intensity. The CCD PSF kernel size was found to increase with increasing intensity and decreasing wavelength.

\end{abstract}

\section{Introduction}\label{sec:intro}

The current cosmological concordance model \citep[e.g.][]{2013arXiv1303.5062P, 2013arXiv1303.5076P, 2013ApJS..208...20B} implies that approximately three quarters of the mass-energy density of the Universe consists of dark energy, while approximately one fifth consists of dark matter. The nature of these constituents is largely unknown. \Euclid \citep[for details, see][and \url{http://www.euclid-ec.org/}]{2011arXiv1110.3193L, 2012SPIE.8442E..0TL, 2014SPIE.9143E..0HL, 2013LRR....16....6A}, the second mission in European Space Agency's Cosmic Vision programme, is designed to make accurate measurements to infer the nature of dark energy, to explore what it is, and to quantify precisely its role in the evolution of the Universe. \Euclid will additionally measure and elucidate the nature of dark matter. If the dark energy is a manifestation of a required modification to general relativity on cosmic scales then \Euclid will also test the validity of modified gravity theories. Until there are high accuracy data to put these new theoretical frameworks to the test, real progress in constraining the nature of the Cosmos will be limited. \Euclid will be one of the most powerful tools in this quest, but only if the systematics can be controlled to an unprecedented accuracy through the combination of technical capability and different cosmological approaches \citep[for a review, see][]{2013LRR....16....6A}.

The main role of the Visible Instrument \citep[VIS;][]{2014SPIE.9143E..0JC} on board the \Euclid satellite is to carry out measurements of the weak gravitational lensing effect by deep imaging of the extra-Galactic sky via the \Euclid Wide Survey \citep{2012SPIE.8442E..0ZA}. To this end, VIS produces images with fine spatial sampling of about 0.1 arc sec over a large field of view, $\sim 0.541$ square degrees, using a camera with 36 4k $\times$ 4k CCD detectors in a $6 \times 6$ mosaic. To provide maximal throughput a single wide bandpass has been adopted with no transmissive elements in the full optical train.

While the standard mode of operation of VIS is simple and in a sense conventional, the level of accuracy that is required of the instrument to measure the weak gravitational lensing effect is exceptional. In particular, good image quality and an extremely high level of knowledge of the Point Spread Function (PSF) are required. The system PSF is a combination of the performance of the optical system, spacecraft pointing accuracy, and the detection system. The detectors and their response to a point source illumination must therefore be known extremely accurately.

Together, the Euclid Consortium, European Space Agency and e2v have designed and manufactured pre-development models of a customised imaging detector for VIS. The new detector is an e2v back-illuminated, 4k $\times$ 4k, 12 micron square pixel CCD designated with a reference number CCD273-84. This device has a higher-responsivity lower-noise amplifier, enhanced red response, parallel charge injection structures and narrower registers which improve low signal charge transfer efficiency \citep[for details, see e.g.][]{CCD273e2v}. The pre-development devices have been studied in detail. For example, pixel-level modelling of charge packet distribution within the CCD273  has been performed \citep{PixelModelling} to address charge transfer inefficiency and correction of it \citep[see e.g.][and references therein]{2014MNRAS.439..887M}. In addition, a front-illuminated pre-development CCD has been tested to observe the CCD PSF relative to signal size using a single-pixel photon transfer curve technique \citep{CCD273FIPSF} and Modulation Transfer Function \citep{CCD273MTFandPSF}. However, a detailed study of the intrinsic PSF performance from charge spreading in the pixel of a back-illuminated CCD273-84 has not been performed and published before. This will be critical for the \Euclid VIS Instrument.

This paper is organised as follows. In Section \ref{sec:data} we describe our laboratory setup used to collected spot projected data. In Section \ref{sec:Bayesian} we briefly review the basics of Bayesian inference and describe the likelihood function, noise model, and the observation model adopted. We also discuss the simplifications and the choice of priors. In Section \ref{sec:results}, we show our results and discuss the implications of the findings. Finally, we summarise our findings and conclude in Section \ref{sec:summary}.

\section{Data} \label{sec:data}

In this Section we detail the laboratory setup that was used to collect spot projection data and discuss what limitations it may set. We also briefly review the data set used in the analysis.

\subsection{Laboratory Setup}\label{ss:labkit}

The spot projection experiments were performed with a back-illuminated CCD273-84, which is based on a previous CCD203 model. A detailed description of the CCD273 is given in e.g. \cite{CCD273e2v} and will not be repeated here. A \Euclid design evaluation model readout electronics (EVM3 ROE) was used to read out the CCD. The analogue ROE chain consists of a correlated double-sampler followed by two successive amplifier stages. The signal is digitised using a low power radiation hard 16-bit Analogue-to-Digital Converter. Digital functionality and clock sequencing is provided via a radiation tolerant FPGA. On-bench power supply units were used to power the electronics. The readout rate was set to 70 kpix/s, while the exposure time was controlled with a programmable shutter. The exposure times were kept as short as possible to minimise the blurring of the optical illumination resulting for example from vibrations.

The testing chamber used for the experiments is split into two sections: a liquid nitrogen cooled cryostat and a vacuum chamber. The vacuum chamber hosts two CCDs, the readout electronics, temperature control system and motorized stages. The temperature of the CCDs and the electronics is controlled independently by two sets of heaters and two sets of temperature sensors. The temperature of the CCDs is set to the \Euclid VIS operational temperature of $-120\deg$C $(\pm 0.1\deg$C$)$ while the temperature of the readout electronics is kept between $0$ and $20\deg$C. The CCDs can be illuminated via a glass window in the chamber wall. 

\begin{figure}
\includegraphics[width=1.\textwidth]{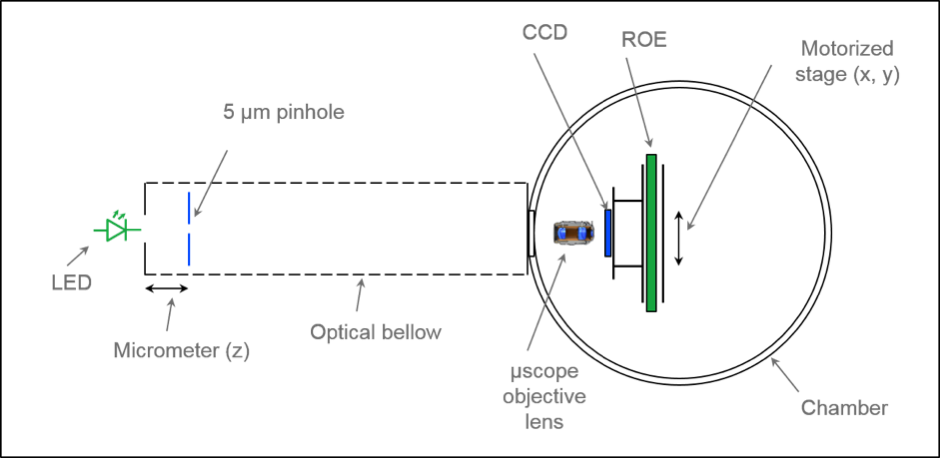}
\caption{Optical Bench during the spot projection setup.}
\label{fig:opticalBench}  
\end{figure}

Figure \ref{fig:opticalBench} shows the optical bench which consists of a LED, $5 \mu$m pinhole mounted on a micrometer stage and a microscope objective lens. LEDs (Roithner Laser Technik) with different wavelengths ranging from 600 to 890nm were used. The diodes were chosen to have narrow spectral widths: FWHMs ranging from 15 to 45nm, allowing near monochromatic illumination at a given wavelength. The $5 \mu$m spot from the high precision pinhole (Thorlabs) was de-magnified onto the CCD surface by a microscope convex lens (Optimus MD Plan) placed at a working distance of 7mm from the CCD surface. The focus of the projection system was adjusted by fine tuning the distance between the pinhole and the CCD with a micrometre screw. The projected spot can be moved across the CCD surface by two micron precision motorized stages moving the CCD.

Given the design of our spot projection setup (Figure \ref{fig:opticalBench}) we can now calculate the theoretical size of the projected spot. The circular aperture diffraction will lead to an Airy pattern with a Full-Width at Half Maximum (FWHM):
\begin{equation}
\mathrm{FWHM_{diffraction}} \sim \frac{1.028\lambda}{\mathrm{NA}} \quad ,
\end{equation}
where $\lambda$ refers to the wavelength and $\mathrm{NA}=0.25$ is the numerical aperture of the system. At the nominal wavelength of 800nm we find $\mathrm{FWHM_{diffraction}} \sim 3.3\mu$m. Our optical system also includes a pinhole with a diameter of 5 microns and a lens with a de-magnification of 10, leading to a geometrical effect of $0.5\mu$m. Because of the relative sizes, we can simply add the two contributions linearly to find a theoretical FWHM of the projected spot, when perfectly in focus, to be $\sim 3.8$ microns at 800nm.

Because the system does not provide automatic focusing, we cannot guarantee that the projected spot was always automatically in the best possible focus. While the micrometre allows focusing the projected spot, the accuracy is limited. In addition, while the vacuum pump is separated from the chamber and connected via a flexible connector, it may lead to very small residual vibrations, which can also lead to a blurring of the projected spot. While such blurring should be small because of short exposure times (see Table \ref{tab:data}), in the model fitting we still need to account for the possible de-focus and any residual blurring that may smooth the projected spot (see Equation \ref{eq:model}). While the mechanical structure holding the CCD is rigid, any vibrations can also lead to slight changes in the exact centre of the projected spot. While none of our data shows that the projected spot would move outside the chosen pixel, it is still possible that the exact centre of the projected spot may move within the pixel. We therefore need to account for the possibility that the projected spot may move slightly from an exposure to exposure. This effect was however confirmed to be relatively small $(< 0.1$ pixels).

\subsection{Available Data}\label{ss:exposures}

Over a two week period several spot projected exposures at different wavelengths and intensity levels were accumulated. Table \ref{tab:data} lists all data used in the following analysis. Several exposures were taken at each wavelength and intensity level to increase the Signal-to-Noise Ratio (SNR) of the data and to allow a rejection of bad data resulting for example from cosmic ray hits. To achieve a high SNR is especially important because the CCD273-84 intrinsic PSF is expected to be narrow and therefore the number of electrons recorded in the pixels neighbouring the illuminated pixel is expected to be low. Furthermore, because the projected spot, while small, is not infinitely small and because of aperture diffraction, some photons illuminate the neighbouring pixels. It is therefore advantageous to have several exposures available to enable more robust separation of the illumination pattern from the actual CCD PSF.

\begin{table}
\caption{A summary of the accumulated spot data used in the analysis.}
\label{tab:data}       
\begin{tabular}{ccrrrr}
\hline\noalign{\smallskip}
Wavelength [nm] & FWHM [$\mu$m] & Peak Intensity $[$ke$^{-}]$ & Exposure Time [s] & Exposures & Peak SNR \\
\noalign{\smallskip}\hline\noalign{\smallskip}
600     & 4.2 &  7.9    &  0.1   &   12   &  89 \\
600     & 4.2 &  21.4    &    0.3  & 12    & 146 \\
600     & 4.2 &  160.2    &   2.5  &  6    & 400 \\
\noalign{\smallskip}
700     & 4.4  &  151.7    &   2.5  &  6    & 389 \\
\noalign{\smallskip}
800     & 4.6 & 7.1    &   0.1  &  10    & 84 \\
800     & 4.6 & 23.4    &  0.4  &   8    & 153 \\
800     & 4.6 & 50.2    &  1.0  &   7    & 224 \\
800     & 4.6 & 82.8    &  1.8  &   6    & 288 \\
800     & 4.6 & 109.3    &  2.6 &   5    &  331 \\
800     & 4.6 & 145.3    &  3.6  &   3    & 381 \\
800     & 4.6 & 157.8    &  4.1  &   5    & 397 \\
800     & 4.6 & 159.4    &  4.2  &   3    & 399 \\
\noalign{\smallskip}
890     & 5.1 & 143.0    &  2.8  &   4    & 378 \\
\noalign{\smallskip}\hline
\end{tabular}
\newline
Note: the SNR is calculated simply by taking into account the Poisson uncertainty in the peak pixel and the readout noise of the detection chain. The FWHM given is the average from data and includes the optical illumination with potential de-focus and the CCD PSF and can therefore differ from the theoretical optical spot size. 
\end{table}

\section{Bayesian Inference}\label{sec:Bayesian}

In this paper we use Bayesian inference allowing us to make probability statements about the interesting model parameters. Before going in to the CCD PSF parameter estimation we discuss our approach and describe our generative model.

\subsection{Background}

In data analysis we are concerned with the question how good a description of the data a given model is. The aim is then to evaluate the posterior probability of the model $M$ given the observed data $D$ in presence of additional ``information'' $I$. The Bayesian paradigm dictates that inference about the model should be based on the probability of it given the data $P(M | D)$. If our model includes a set of parameters, denoted with $\bar{\theta}$, whose values we want to estimate from our data $D$, we can write Bayes' theorem \citep{Bayes01011763} as follows 
\begin{equation}\label{eq:Bayes}
P(M, \bar{\theta} | D, I) = \frac{P(D | M, \bar{\theta}, I)P(M, \bar{\theta} | I)}{P(D | I)} \quad .
\end{equation}
In this case we have included additional information $I$ to the analysis by including probability distributions for all model parameters $P(M, \bar{\theta} | I)$ as priors.

To search for the optimal\footnote{The ``optimal'' can be in this context understood as a point estimator with the highest posterior probability. However, in reality we are more interested in the credible intervals of the model parameters.} model $M$ parameters $\bar{\theta}$, which maximise the posterior probability density function $P(M, \bar{\theta} | D, I)$, we must solve Equation \ref{eq:Bayes}. Because in practice it is often very difficult to solve for the ``evidence'' $P(D|I)$, and since this is only a normalisation factor, we are more interested in sampling from the product of the likelihood function $P(D | M, \bar{\theta}, I)$ and the priors $P(M, \bar{\theta} | I)$. The choice of the probability functions for the priors is subjective. Even when we choose the most uninformative priors the posterior estimate is still dependent on this choice. Care must therefore be exercised when choosing the priors. In the following analyses we in general choose flat or non-informative priors when no good knowledge of the actual parameter value is available. We do however limit the parameter values to a realistic range to exclude unphysical solutions and to speed up the sampling of the posterior.

\subsection{Noise Model and the Likelihood Function}\label{ss:noiseModel}

A noise model for CCD data should include photon counting statistics describing the detection process and electronics noise describing the read out process. The detection process can be describe by Poisson statistics $\mathcal{P}(\mu)$, where $\mu$ is the expected number of photoelectrons, while the reading out process by a Normal distribution with zero mean: $\mathcal{N}(0, \sigma^{2})$, where $\sigma^{2}$ is the variance that can be associated with electronics readout noise. Thus, we can assume that datum $D$ per pixel $(i, j)$, i.e. the number of electrons recorded in that pixel, is a combination of two independent terms: $D_{i, j} = \mathcal{P}_{i, j} + \mathcal{N}_{i, j}$, where $\mathcal{P}_{i, j}$ is a discrete Poisson distribution and a function of the source and $\mathcal{N}_{i, j}$ is a continuous Normal (Gaussian) distribution.

Because the probability distribution of the sum of two (or more) independent random variables is the convolution of their individual distributions, the probability of the data $D$ given a model $M$ i.e. the likelihood, for each pixel $(i, j)$, can be written as
\begin{equation}\label{eq:dataPoisNormal}
P( D_{i, j} | M_{i, j}, \bar{\theta}, I) = P_{\mathcal{P}}(\mathcal{P}_{i, j} | M_{i, j}, \bar{\theta}, I) \Conv P_{\mathcal{N}}(\mathcal{N}_{i, j} | M_{i, j}, \bar{\theta}, I) \quad ,
\end{equation}
where $\Conv$ is a convolution. Using Bayes' theorem, substituting  Equation \ref{eq:dataPoisNormal} to \ref{eq:Bayes}, we can now write the full posterior probability as
\begin{equation}
P(M_{i, j}, \bar{\theta} | D_{i, j}, I) = \frac{\left [ P_{\mathcal{P}}(\mathcal{P}_{i, j} | M_{i, j}, \bar{\theta}, I) \Conv P_{\mathcal{N}}(\mathcal{N}_{i, j} | M_{i, j}, \bar{\theta}, I) \right ]P(M_{i, j}, \bar{\theta} | I)}{P_{\mathcal{P}}(\mathcal{P}_{i, j} | I) \Conv P_{\mathcal{N}}(\mathcal{N}_{i, j} | I)} \quad .
\end{equation}
This equation describes the posterior probability of the model $(M_{i, j}, \bar{\theta})$ given data $(D_{i, j})$ and information $I$ for a pixel $(i, j)$. 

However, as already mentioned, in practice we often do not need to solve for the evidence $P(D | I)$. We can therefore concentrate on the likelihood function $P(D | M)$ and the priors $P(M| I)$. Because in our case $P_{\mathcal{P}}$ is a Poisson and $P_{\mathcal{N}}$ is a Normal distribution, a complication arises as $P_{\mathcal{P}}$ is discrete while $P_{\mathcal{N}}$ is continuous. Nonetheless, from the definition of a convolution we can rewrite the likelihood, Equation \ref{eq:dataPoisNormal}, as
\begin{eqnarray}
P(D_{i, j} | M_{i, j}) & = & \int \mathrm{d}D' P_{\mathcal{P}}(D' | M_{i,j})P_{\mathcal{N}}(D_{i, j} - D' | M_{i, j}) \\
& = & \int \mathrm{d}D' \left [ \frac{\exp^{-D'}D'^{M_{i, j}}}{M_{i, j}!} \right ] \left [ \frac{1}{\sqrt{2 \pi \sigma^{2}_{i, j}}} \exp^{-\frac{(D_{i, j} - D' - M_{i, j})^{2}}{2\sigma^{2}_{i, j}}} \right ] \label{eq:xx} ,
\end{eqnarray}
where $\sigma^{2}_{i, j}$ is the variance of the Normal distribution describing the read out noise and $M_{i, j}$ is a discrete model. Note that above we have not written the model parameters $\bar{\theta}$ or the information $I$ explicitly for clarity.

Assuming that each pixel is uncorrelated the total probability over the image pixels is just then the product of the individual pixel probabilities
\begin{equation}\label{eq:product}
P(D | M) = \prod_{i, j}^{N_{\textrm{{pix}}}} P(D_{i, j} | M_{i, j}) \quad .
\end{equation}
Substituting Equation \ref{eq:xx} in \ref{eq:product}, the likelihood function can then be written as
\begin{equation}\label{eq:fullLikelihoodFunction}
P(D | M, \bar{\theta}) = \prod_{i, j}^{N_{\textrm{pix}}} \left \{ \int \mathrm{d}D' \left [ \frac{\exp^{-D'}D'^{M_{i, j}}}{M_{i, j}!} \right ] \left [ \frac{1}{\sqrt{2 \pi \sigma^{2}_{i, j}}} \exp^{-\frac{(D_{i, j} - D' - M_{i, j})^{2}}{2\sigma^{2}_{i, j}}} \right ] \right \} . 
\end{equation}

However, for practical reasons we take advantage of the fact that most of the projected spots are very bright, containing from $10^{4}$ to $10^{5}$ photoelectrons in the peak pixel and $10^{2}$ to $10^{4}$ in the neighbouring pixels. At such high flux levels Poisson distribution tends to a Gaussian (the central limit theorem). Now, instead of having a sum of Poisson and Normal distribution, we have a sum of two independent Normal distributions with different means and variances: 
\begin{equation}
D_{i, j} = \mathcal{N}_{i, j}(\mu,\, \sigma_{1}^{2}) + \mathcal{N}_{i, j}(0,\, \sigma_{2}^{2}) \quad .
\end{equation}
Because a convolution of two Gaussians is a Gaussian we can describe the $D_{i, j}$ simply with a single Normal distribution:
\begin{equation}
D_{i, j} = \mathcal{N}_{i, j}(\mu, \, \sigma^{2}) \quad ,
\end{equation}
where the variance $\sigma$ is the sum of the variances of the two independent Gaussian distributions and $\mu$ refers to the number of electrons in a pixel $(i, j)$. With this assumption, we can simplify the likelihood function, Equation \ref{eq:fullLikelihoodFunction}, and write
\begin{equation}\label{eq:gaussianLikelihood}
P(D | M, \bar{\theta}) = \prod_{i, j}^{N_{\textrm{pix}}} \frac{1}{\sqrt{2 \pi \sigma^{2}_{i, j}}} \exp^{-\frac{(D_{i, j} - M_{i, j})^{2}}{2\sigma^{2}_{i, j}}}  \quad .
\end{equation}

In the inference we are searching for parameters $\bar{\theta}$ that maximise the product of the likelihood with the priors. However, in practise, because these probabilities can become very small, we instead maximise the sum of the log-likelihood $\mathcal{L} = \log [P(D | M, \bar{\theta})]$ and log-priors. Taking the logarithm of Equation \ref{eq:gaussianLikelihood}, we finally arrive to the log-likelihood
\begin{equation}
\mathcal{L} = - \frac{N_{\textrm{pix}}}{2}\log(2\pi) - N_{\textrm{pix}} \sum_{i, j}^{N_{\textrm{pix}}} \log(\sigma_{i, j}) - 
\sum_{i, j}^{N_{\textrm{pix}}} \frac{\left(D_{i, j} - M_{i, j}\right)^{2}}{2\sigma_{i, j}^{2}} \quad .
\end{equation}
In the sampling process, we can omit the constant terms and simply write
\begin{equation}\label{eq:simplifiedLikelihood}
\mathcal{L} = - \frac{1}{2} \sum_{i, j}^{N_{\textrm{pix}}} \left [ \frac{\left(D_{i, j} - M_{i, j}\right)^{2}}{\sigma_{i, j}^{2}} \right ] \quad ,
\end{equation}
where $\sigma_{i, j}$ refers to a total CCD noise model. This is the log-likelihood function that we will use in the following analysis.

The total CCD noise model $\sigma_{i, j}$ (in digital numbers; DNs) for each pixel $i, j$ can be written as
\begin{equation}\label{eq:CCDnoise}
\sigma_{i, j} = \sqrt{\frac{M_{i, j} - B}{g} + \left ( \frac{\textrm{RN}}{g} \right )^{2}} \quad ,
\end{equation}
where $B$ refers to an ADC offset level (bias), $g$ denotes a gain factor, and $\textrm{RN}$ refers to the readout noise. For our setup, we have inferred the gain from photon transfer curve data an assume it to be constant over the area that a projected spot covers. The ADC offset level is derived from the pre-scan region, while the readout noise is inferred as the standard deviation of the pixels on the same row as the projected spot, but not illuminated by a light source or covered by cosmic rays or dead/hot pixels. It should be noted that the results are not sensitive to a small inaccuracies in these terms and therefore we chose not to model them. However, for a less stable system all the parameters in Equation \ref{eq:CCDnoise} could be different for each pixel $(i, j)$.

\subsection{Observation Model}\label{ss:ObseravationModel}

In this Section we discuss the modelling of the laboratory setup and the charge diffusion to infer the CCD PSF from spot projected data. We start by describing our model $M_{i, j}$. To model the observed pixel data $\{D_{i, j}\}$ at pixel $(i, j)$, we assume that the data are generated by a process described by the following equation:
\begin{equation}\label{eq:model}
M_{i, j} = \left \{ \left [ A_{\textrm{pinhole}} \Conv G^{\textrm{sym}}_{\textrm{focus}} \right ] \Conv G_{\textrm{CCD}} \right \}_{i, j} \quad .
\end{equation}
In Equation \ref{eq:model} $A_{\textrm{pinhole}} = A(I, x, y, r, \lambda)$ refers to an Airy disc that describes the light exiting the pinhole (round aperture diffraction), $G^{\textrm{sym}}_{\textrm{focus}} = G(x, y, \sigma_{xy}, \lambda)$ refers to a circular symmetric smoothing generated by a slight out-of-focus position and potential blurring, $G_{\textrm{CCD}} = G'(x', y', \sigma_{x}, \sigma_{y}, \lambda)$ describes the CCD PSF approximated with a non-symmetric two-dimensional Gaussian, and finally $\Conv$ denotes a convolution. It is assumed that both the Airy disc, with an intensity $I$ (note that here we use $I$ to refer to intensity not to information), and radius $r$, and the de-focus smoothing kernel are located at the same position $(x, y)$, while the CCD PSF is centred on the central pixel $(x', y')$. Finally the model is placed on the CCD pixel grid $(i, j)$.

Because our laboratory setup generates near monochromatic light, we can omit the wavelength dependency, denoted with $\lambda$, from the terms appearing in Equation \ref{eq:model}. We are therefore left with the following set of model parameters $\bar{\theta} = \{I, x, y, r, \sigma_{xy}, x', y', \sigma_{x}, \sigma_{y} \}$. The inclusion of the CCD PSF anisotropy, $\sigma_{x}$ and $\sigma_{y}$, allows the model to create elliptical profiles, which is important for this study.

Because the intensity $I$, or amplitude, of the Airy disc at location $(x, y)$ depends on the distance of that location from a point of reference (chosen to be the CCD PSF position $x', y'$) and the Bessel function of the first kind of order one, it was found to be sensitive to the derived position $(x, y)$. Moreover, we are not interested in the exact value of the Airy disc, because the observed data are pixelised. We therefore adopt a meta-parameter $m$ which links the counts in the peak pixel to the intensity of the Airy disc. It should be noted that to estimate the amplitude using the peak pixel value depends also on the radius $r$ of the Airy disc. In our chosen model, while the radius is a free parameter, it is in practice connected to the amplitude via the meta-parameter $m$, which can be defined as:
\begin{equation}
m(p, x, y, r, x', y') = p \left[\frac{2 J_1\left(\frac{\pi d}{r/R_z}\right)}{\frac{\pi d}{r/R_z}}\right]^{-2} \quad ,
\end{equation}
where $p$ refers to the number of electrons in the peak pixel, $J_{1}$ is the first order Bessel function of the first kind, $d$ is a radial distance from the maximum of the Airy disc $d = \sqrt{(x - x')^{2} + (y - y')^{2}}$, and $R_{z} \sim 1.22$ (the first zero of $J_{1}\pi^{-1}$). We can therefore now write the description of the Airy disc as $A_{\textrm{pinhole}} = A(m)$.

\begin{figure}
\includegraphics[width=1.\textwidth]{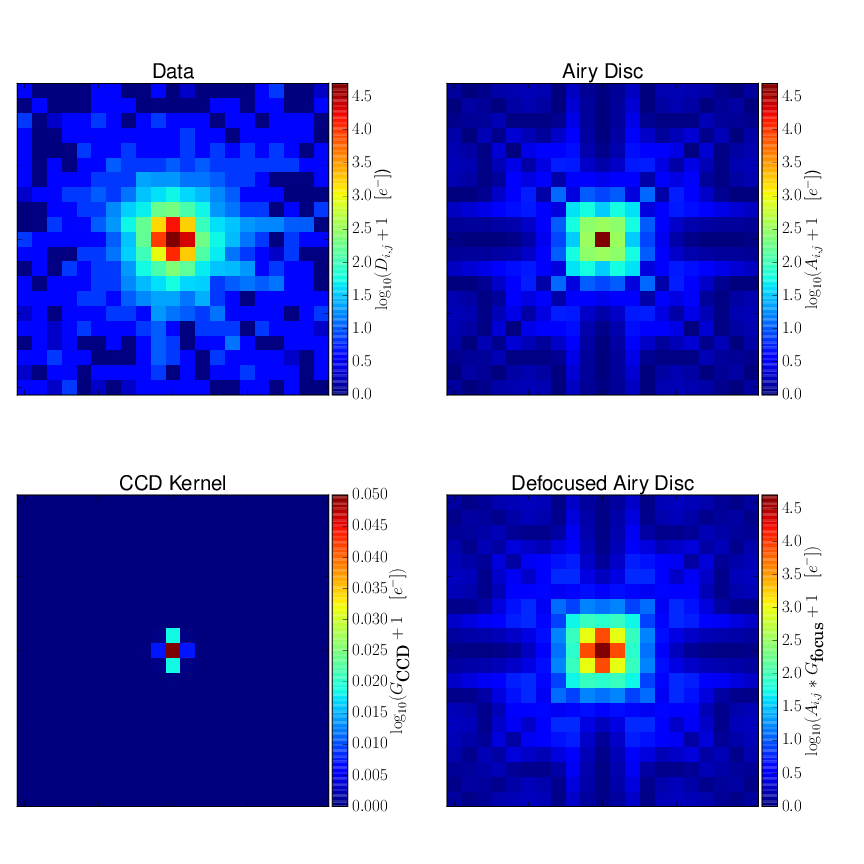}
\vspace{-8mm}
\caption{Example of the model components and data at 800nm. Note that no model has been fit to the data, but the properties of the model components were chosen by hand to illustrate the different components of the model. Note that the Airy disc looks like a square because of the sampling is on the nominal CCD273 pixel scale.}
\label{fig:modelExample}  
\end{figure}

Figure \ref{fig:modelExample} shows an example of the model components given in Equation \ref{eq:model} together with an example of the spot projected data. The properties of the model, such as intensity and sizes, are chosen by hand and to illustrate the different components of the model and are hence not fit to the data shown. It is evident from the Figure that to obtain a robust estimate for the properties of the CCD PSF the anisotropy of the charge leakage is of key importance. In addition, if the Airy disc is slightly offset from the exact pixel centroid, as is the case for most laboratory data, the degeneracy between the defocus and the CCD PSF can be broken. It is these two aspects that allow us to recover the properties of the CCD PSF, as will be shown next.

\subsection{Testing Parameter Recovery with Simulated Data}\label{ss:simulatedData}

To test our model, parameter recovery and the assumptions adopted, we generated a set of simulated data. To mimic the observed spot projected data, we adopted the following procedure:
\begin{enumerate}
\item Generate a high-resolution description of the pinhole with an Airy disc,
\item Apply defocus,
\item Apply CCD diffusion kernel,
\item Pixelise the data to the native CCD pixel grid,
\item Add Poisson noise,
\item Add an ADC offset (bias) level,
\item Add readout noise, assumed to be drawn from a Gaussian with a standard deviation of 4.5e$^{-}$ (as specified for \Euclid VIS),
\item Convert the electrons to digital numbers by applying a gain (3.1 e$^{-}$/DN, as specified for \Euclid VIS), and
\item Store the generated image as a 16bit integer image array.
\end{enumerate}

Data simulated with the above procedure should produce images that are closely matched to the real data and indeed, when the parameters, such as the intensity and radius of the Airy disc and de-focus and noise parameters, are correctly chosen, the simulated data are a close proxy to the observed spot projected images. Because we have to choose the model parameters when simulating the data, we know the input model and we can try to recover them from the simulated data $D$. This allows us to estimate how to sample the posterior for accurate recovery.

To a first instance we simulated five exposures of a well-exposed spot with $\sim 1.5 \times 10^{5}$ electrons in the peak pixel. This is well within the capabilities of a CCD273, which has a full-well capacity of $\sim 2 \times 10^{5}$e$^{-}$. The remaining model parameters are chosen to mimic the laboratory setup. We set the radius of the Airy disc to 0.47 pixels, defocus $\sigma_{xy} = 0.41$, and the CCD PSF kernel, which we are trying to recover, to $(\sigma_{x}, \sigma_{y}) = (0.291, 0.335)$ pixels. The centroid of the Airy disc and hence the de-focus kernel was varied within $\pm 0.1$ pixels between different images, as this was assumed to resemble the small variations in the experimental setup.

\begin{figure}
\includegraphics[width=1.\textwidth]{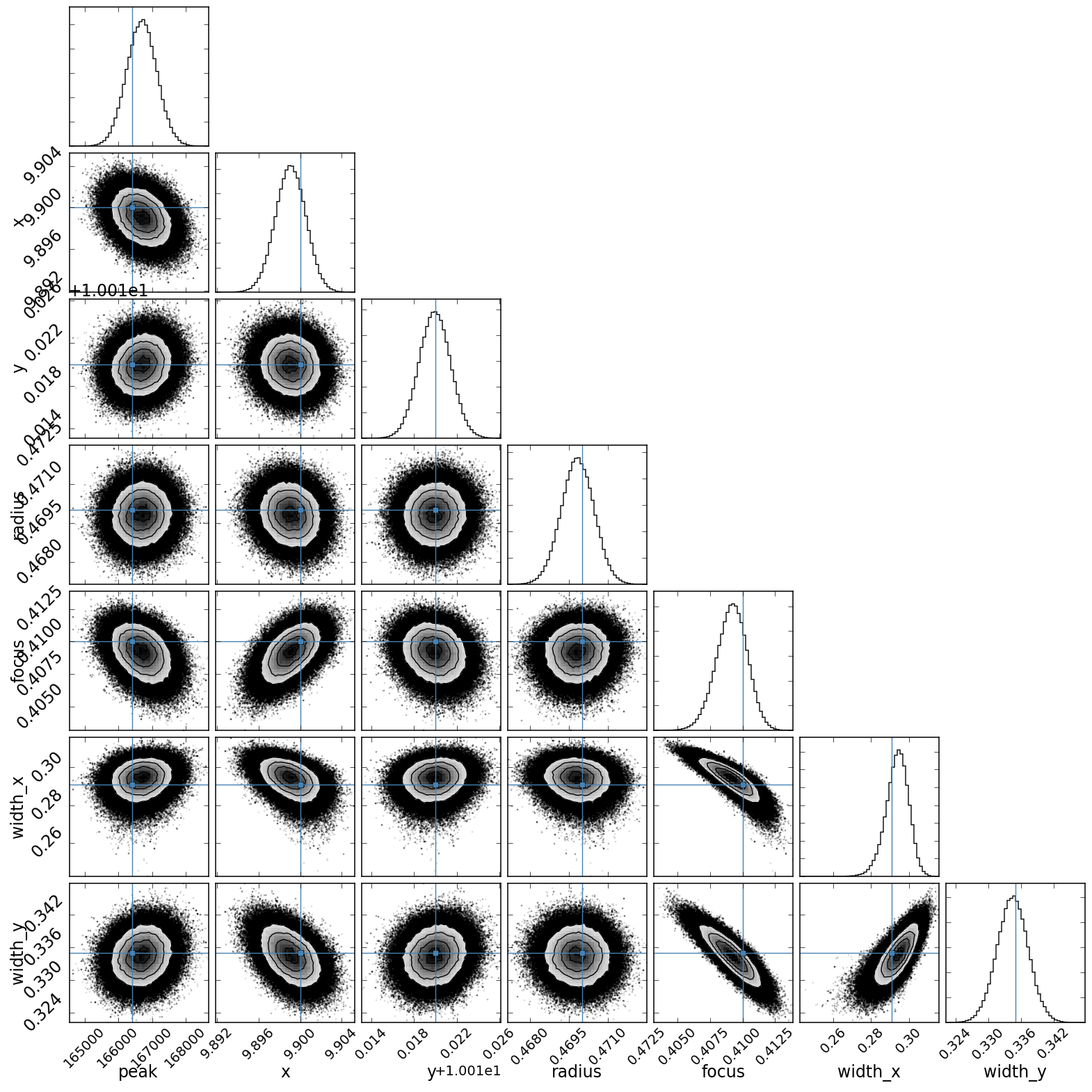}
\caption{The one and two dimensional projections of the posterior probability distributions of model parameters. The Figure shows the marginalized distribution for each parameter independently in the histograms along the diagonal and then the marginalized two dimensional distributions in the other panels. The input parameters for the simulation are given in blue. The contours shown are 0.5, 1, 1.5, and $2\sigma$. Generated with a code given in \cite{Foreman-Mackey:11020}.}
\label{fig:singleFitSimulatedTriangle}  
\end{figure}

We use Affine Invariant Markov Chain Monte Carlo Ensemble sampler \textit{emcee} \citep[version 2.1.0;][]{2013PASP..125..306F} to map the posterior given the log-likelihood function of Equation \ref{eq:simplifiedLikelihood} and our chosen priors. We start by sampling a probability density function for the model parameters, discussed in Section \ref{ss:ObseravationModel}, for each individual exposure separately. An example of the parameter recovery in a single representative case is given in Figure \ref{fig:singleFitSimulatedTriangle}. It is evident from this Figure that some parameters (de-focus and the CCD PSF widths) are degenerate and therefore more difficult to recover. In general, however, the parameter recovery is relatively good, as truth, indicated by the intersections of the blue lines, is within reasonable credible intervals. The most interesting parameters, the size of the CCD kernel (width\_x and width\_y), are reasonably well recovered.

It is however possible to have more stringent constraints on the CCD PSF recovery, if we assume that it does not vary from exposure to exposure, but allow the spot to have moved slightly. In such a case we can do a joint fit to the data, by allowing the Airy disc position to change from image to image, but requiring the CCD PSF parameters to be fixed. As the LEDs providing the illumination in our laboratory setup are relatively stable (after stabilisation period the irradiance varies $< 0.1$ per cent), we can further assume that the intensity of the illumination stays fixed over the experiment. Furthermore, we can assume that the defocus is also stable over short time periods it takes to collect from a few to a few tens of exposures. We can therefore fix (between images) several parameters in the joint fit to reduce the dimensionality of the problem, which allows improved constraints to be drawn.

\begin{figure}
\includegraphics[width=1.\textwidth]{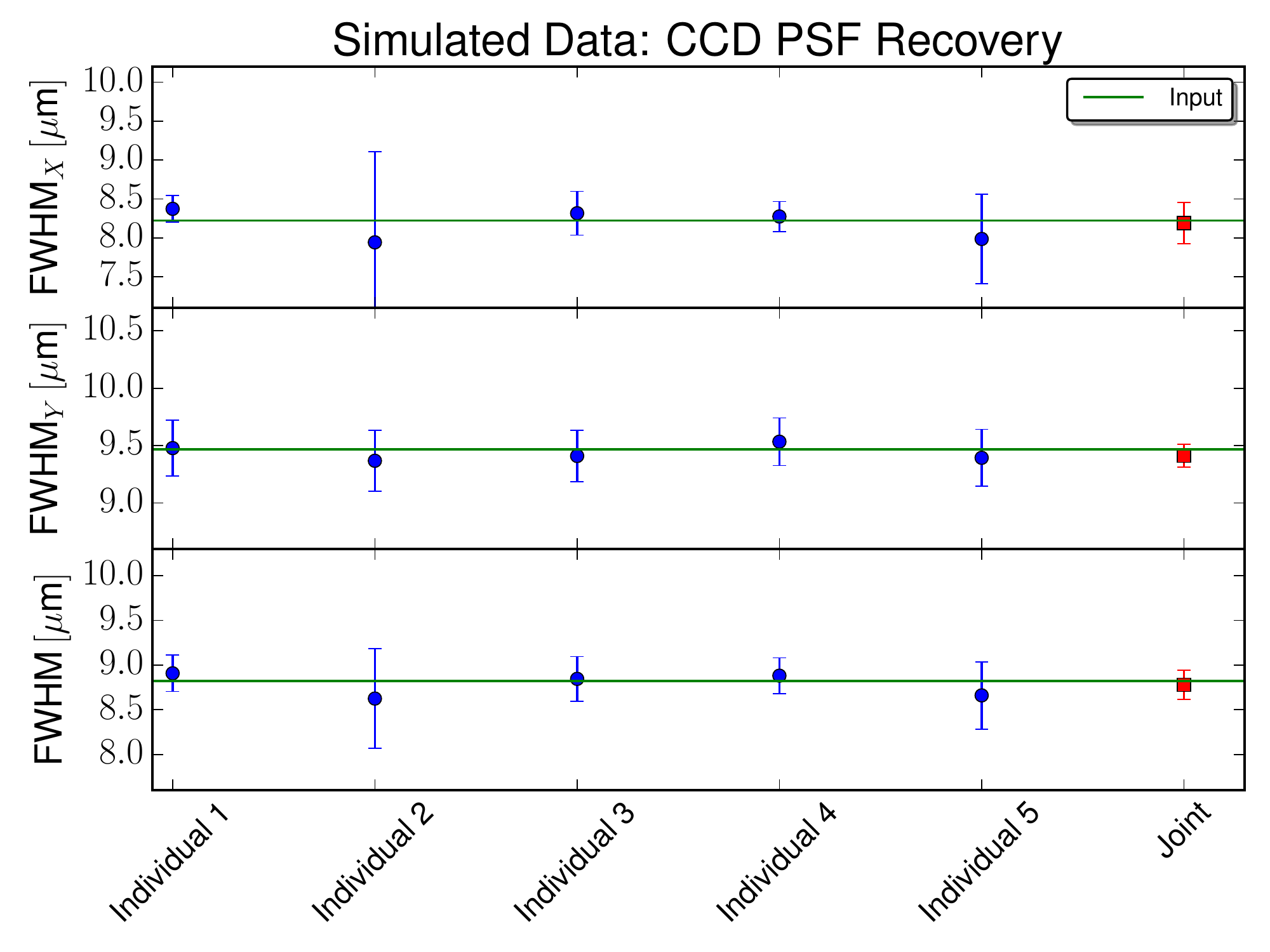}
\caption{Recovering the CCD PSF kernel size from either individually fitting each exposure or jointly. The input true values of the CCD PSF size are given by the horizontal green lines. The uncertainties shown are $3\sigma$ estimates from the MCMC chains.}
\label{fig:SimulatedVsTruth}  
\end{figure}

If we model the CCD PSF jointly from all suitable exposures we obtain an improved constraint for the parameters of interest. Figure \ref{fig:SimulatedVsTruth} shows the size estimates for the CCD PSF kernel, if they are derived from five exposures independently or adopting joint modelling. The input parameter values that were used to simulate the data are shown with green horizontal lines. While the individual estimates are in most cases close to this line, the joint modelling provides a more accurate estimate with smaller uncertainties (here given as $3\sigma$ limits).

\begin{figure}
\includegraphics[width=1.\textwidth]{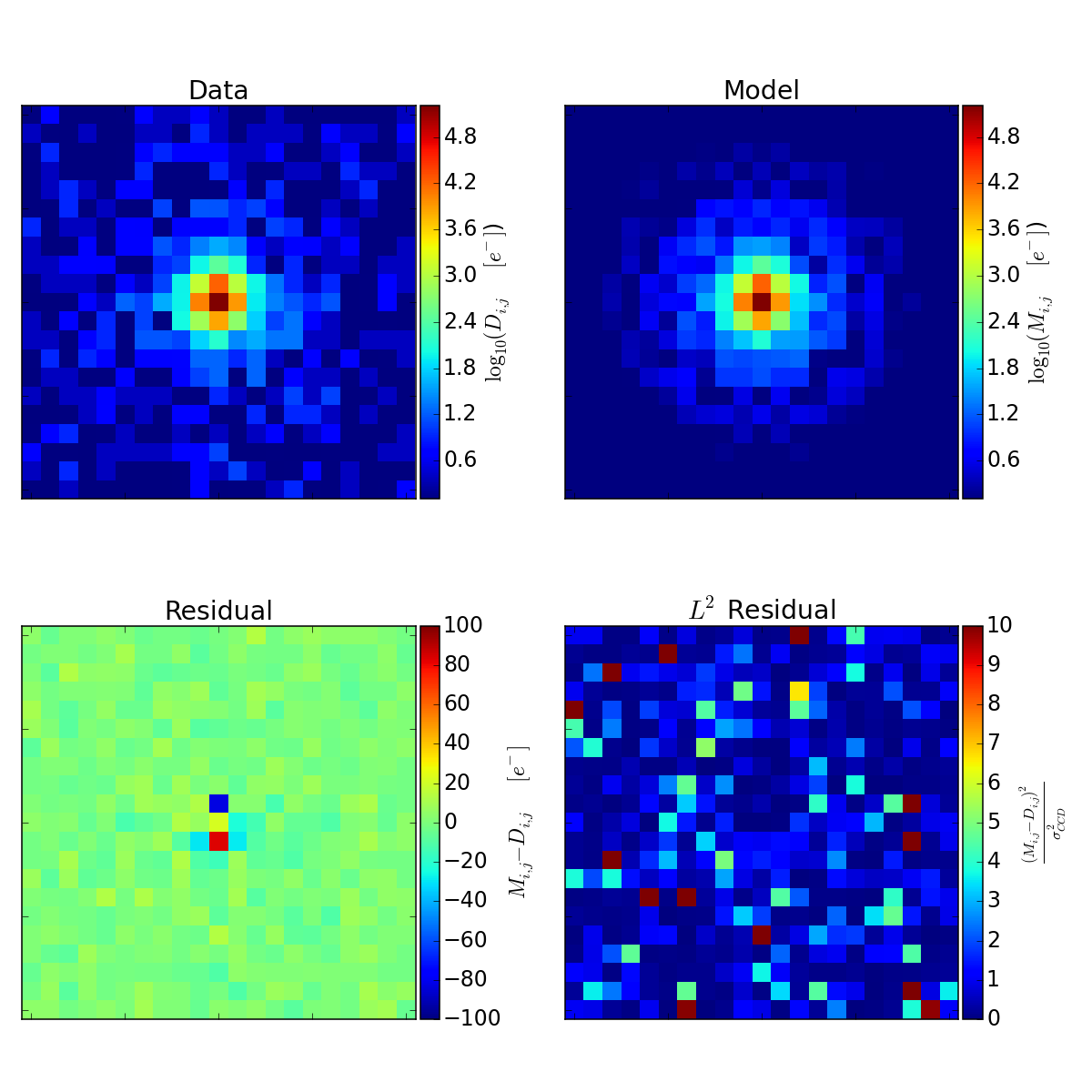}
\caption{Comparison of the derived model $M_{i, j}$ against simulated data $D_{i, j}$. The example is representative of the other fits.}
\label{fig:SimulatedResiduals}  
\end{figure}

To summarise, with simulated data we have shown that even with the rather small number of pixels the spots cover, our relatively complicated model, with potentially some inherent degeneracies\footnote{Because a convolution of two Gaussians is a Gaussian, the defocus and the CCD kernel are degenerate if the projected spot is exactly centred on the pixel. In practice this is never the case.}, can recover the true underlying parameter values. This is possible because the information content in the data is sufficient. Nonetheless, while Figures \ref{fig:singleFitSimulatedTriangle} and \ref{fig:SimulatedVsTruth} show that we can successfully recover the underlying truth, it should be kept in mind that the tests are simplistic. Because we have generated the simulated data with the same model that is used in the parameter estimation, we therefore implicitly assume that we know perfectly the generative process producing the spot projected data. Because this may not be the case for the real laboratory data, we tested several different models before settling to the one described above (Equation \ref{eq:model}) as it was found to describe the observed data most closely. The other models tested included a simplified version of Equation \ref{eq:model} where we did not apply the de-focus term. However, this led to large residuals and artificially large CCD PSF and was therefore considered not to reflect reality. We also tried to decouple the de-focus kernel location from the Airy disc location. The fitting and marginalised posteriors, however, implied that these two should have the same centre. This intuitive assumption was therefore confirmed with the observed data. While a true Bayesian would use all possible models and then Bayesian model average \citep[see e.g.][and references therein]{hoeting1999} to help account for the uncertainty inherent in the model selection process, this is beyond the scope of the current work. We therefore conclude that the model given by Equation \ref{eq:model} is a fair approximation of the process generating the observed spot projected data and should allow the recovery of the intrinsic CCD PSF.

\section{Results: CCD273 PSF Characteristics} \label{sec:results}

\subsection{Size and Ellipticity}

The key CCD characteristics include the size and ellipticity of the PSF. Depending on the assumptions the size of the CCD PSF can be parametrised in several ways. For $\textit{Euclid}$ two different size measures have been chosen: 1) the Full Width at Half Maximum (FWHM) of a Gaussian, and 2) the weighted size $R^{2}$. The former metric is more conventional in Astronomy, while the latter is useful for example in weak gravitational lensing where the PSF wings are important \citep[e.g.][]{2013MNRAS.429..661M}. In the following we report the CCD273-84 PSF size using both metrics. Because we are interested in the shape measurements of galaxies, we are also interested in the ellipticity $e$ of the PSF. For two-dimensional Gaussian we can write these three quantities as follows:
\begin{eqnarray}
\textrm{FWHM} & = & 2\sqrt{2\sigma_{x}\sigma_{y}2\ln{2} }\label{eq:FWHM} \\
R^{2} & = &  \sigma_{x}^{2} + \sigma_{y}^{2}\\
e & = & \left | \frac{\sigma_{x}^{2} - \sigma_{y}^{2}}{\sigma_{x}^{2} + \sigma_{y}^{2}} \right | \quad ,
\end{eqnarray}
where $\sigma_{x}$ and $\sigma_{y}$ are the widths of the Gaussian in $x$ (row/serial) and $y$ (column/parallel) direction, respectively.

\begin{figure}
\includegraphics[width=1.\textwidth]{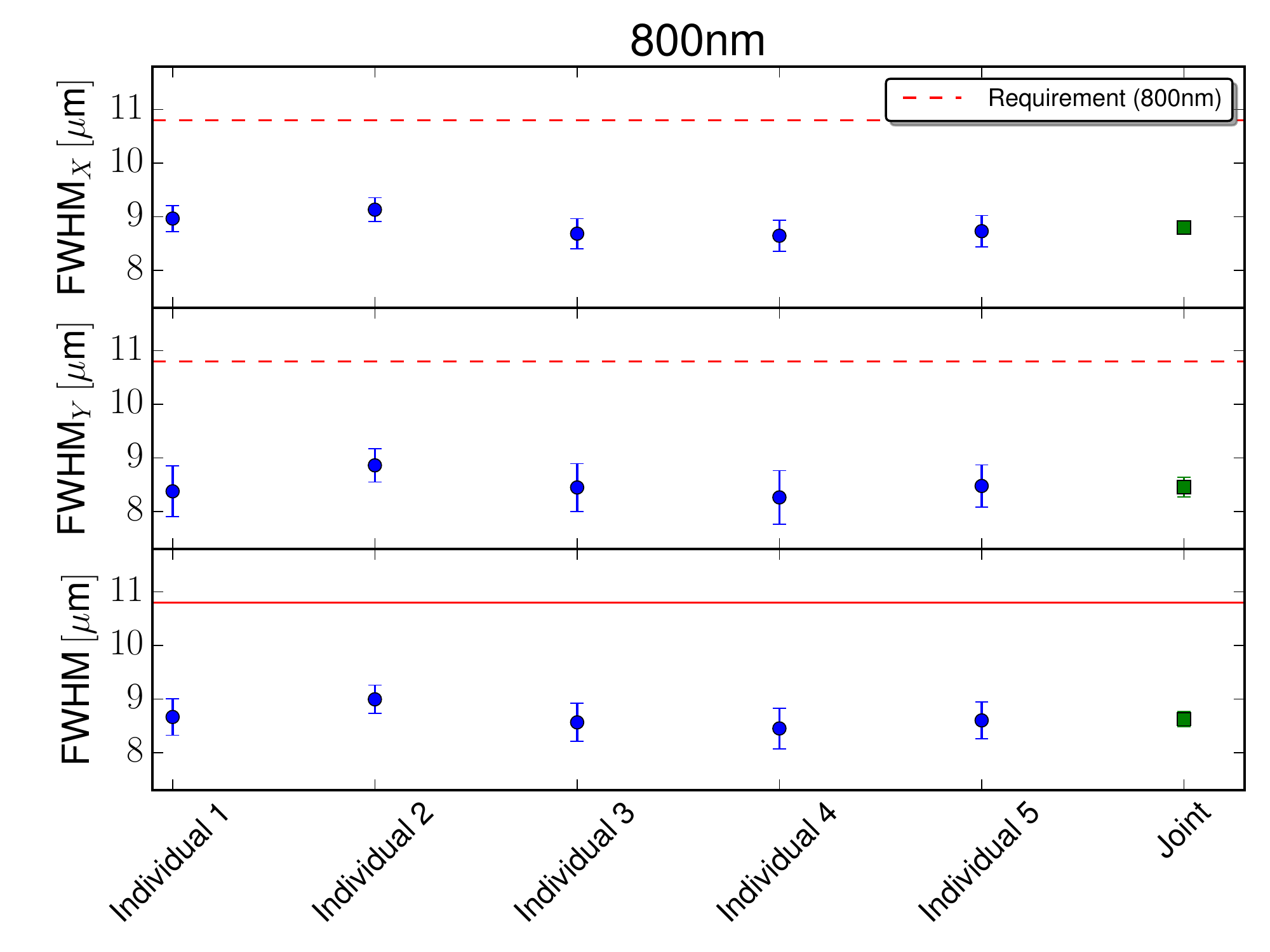}
\caption{CCD273 PSF size in parallel (y) and serial (x) direction in microns measured at 800nm from exposures reaching 150ke$^{-}$ or $\sim 75$ per cent of the full-well capacity. The uncertainties shown are $3\sigma$ estimates from the MCMC chains. The \Euclid VIS requirement has been plotted with red.}
\label{fig:800nmFWHM}  
\end{figure}

Figure \ref{fig:800nmFWHM} shows the FWHM of the CCD273-84 PSF measured at 800nm from data exposed at $\sim 75$ per cent of the full-well capacity (i.e. $\sim 150$ke$^{-}$). The FWHM is reported in both serial (x) and parallel (y) direction and a single size measure is given as the geometric mean of the two (Equation \ref{eq:FWHM}). The Figure shows that the size of the CCD273 PSF at 800nm is $\sim 8.8$microns and well within the $\textit{Euclid}$ VIS requirement of 10.8 microns. Interestingly, the Figure shows that the size of the CCD273-84 PSF is very similar in both serial and parallel directions, leading to a very low PSF ellipticity from the detector system.

\begin{figure}
\includegraphics[width=1.\textwidth]{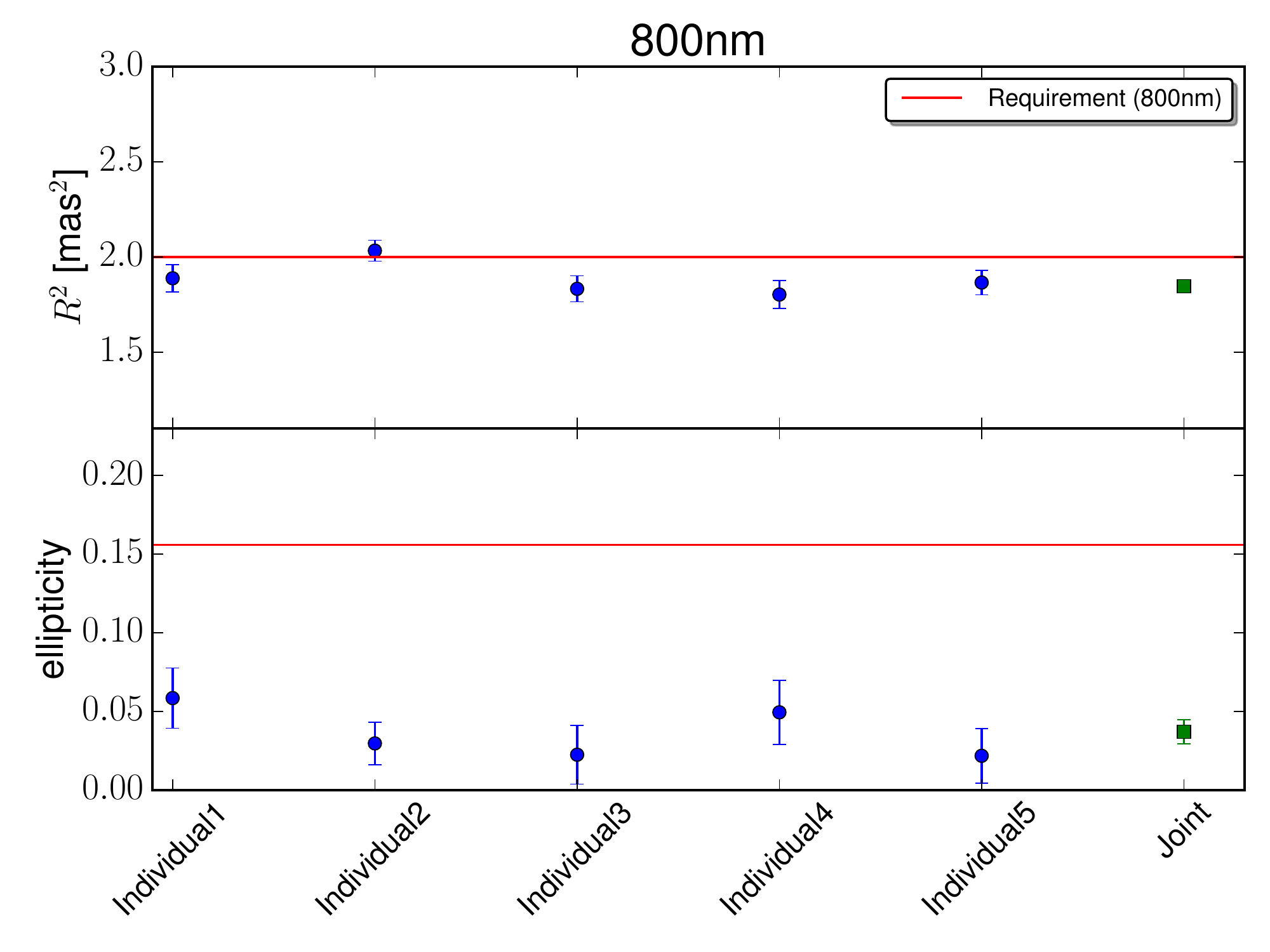}
\caption{CCD273 PSF weighted size $R^{2}$ and ellipticity measured at 800nm from exposures reaching 150ke$^{-}$ or $\sim 75$ per cent of the full-well capacity. The uncertainties shown are $3\sigma$ estimates from the MCMC chains. The \Euclid VIS requirement has been plotted with red.}
\label{fig:800nmR2ell}  
\end{figure}

Figure \ref{fig:800nmR2ell} shows the CCD273-84 PSF weighted size $R^{2}$ and ellipticity at 800nm. The weighted size of the PSF is just under 2 milliarcseconds squared, but within the requirement. The lower panel of the Figure shows that the CCD273-84 PSF is very round, the ellipticity being $\sim 5$ per cent and well within the requirement of 15.6 per cent. This is especially good for a camera build for a weak gravitational lensing mission for which more circularly symmetric PSF leads to improved overall performance \citep{2013MNRAS.431.3103C}. It should be noted however that this does not include the contribution from the telescope system.

\begin{figure}
\includegraphics[width=1.\textwidth]{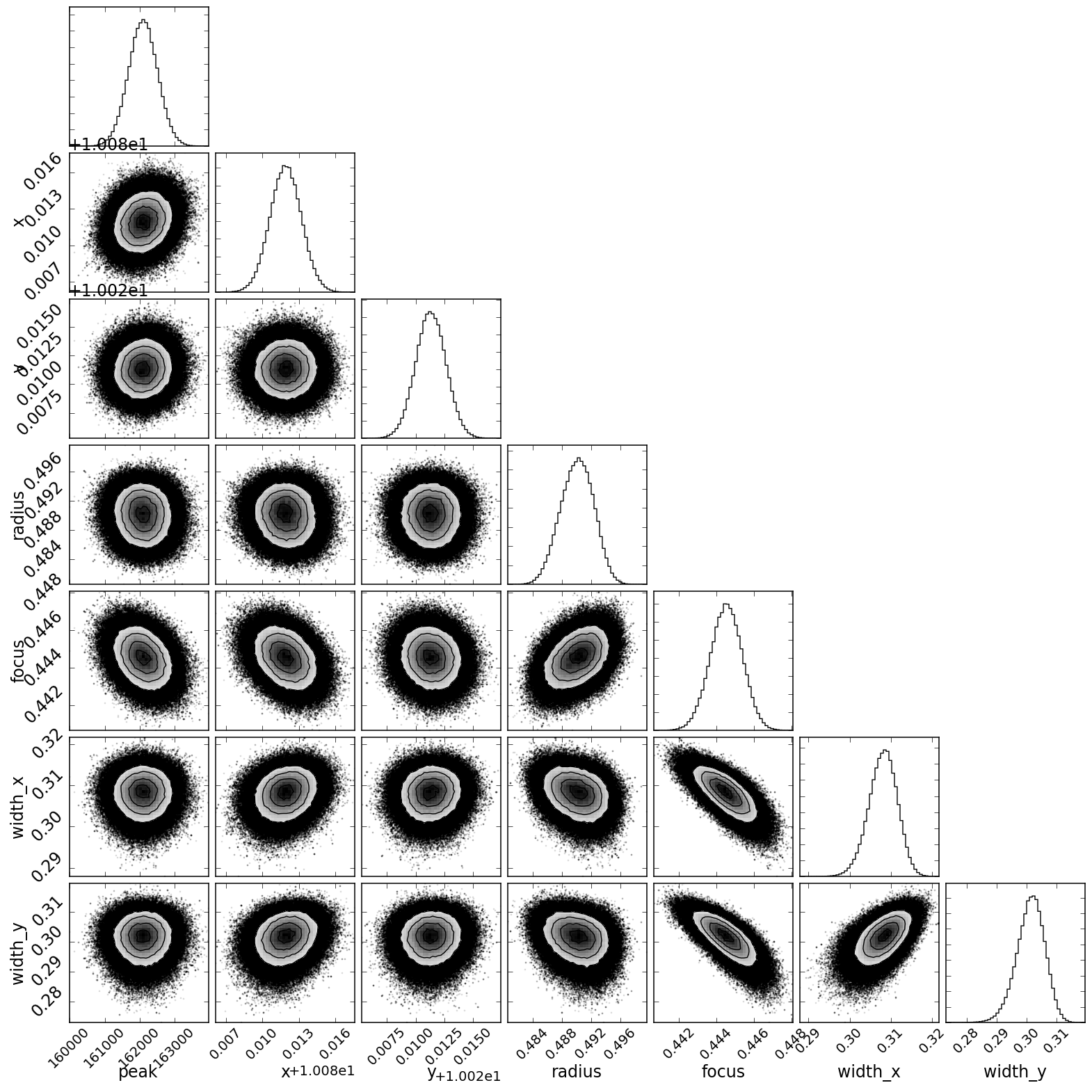}
\caption{The one and two dimensional projections of the posterior probability distributions of model parameters. The Figure shows the marginalized distribution for each parameter independently in the histograms along the diagonal and then the marginalized two dimensional distributions in the other panels. The contours shown are 0.5, 1, 1.5, and $2\sigma$.}
\label{fig:800nmTriangle}  
\end{figure}

Figure \ref{fig:800nmTriangle} shows the one and two dimensional projections of the posterior probability distributions of model parameters when fitted to spot projected data collected at 800nm. The projected distributions show that the parameters are most likely rather well recovered, albeit there is a degeneracy between the de-focus and the size of the CCD PSF. The Figure is comparable to the results from simulations (Fig. \ref{fig:singleFitSimulatedTriangle}). We can therefore conclude that the CCD273-84 PSF characteristics have been recovered accurately.

\subsection{Wavelength Dependency}\label{ss:wavelength}

We have also characterised the CCD273-84 PSF size as a function of wavelength. The detector PSF size will change as a function of wavelength because blue photons are converted to photoelectrons closer to the back surface than red photons. We can therefore expect that the CCD PSF is larger at bluer wavelengths because photoelectrons have longer path to travel in a weaker electric field leading to a larger probability of migrating to the neighbouring pixels. 

\begin{figure}
\includegraphics[width=1.\textwidth]{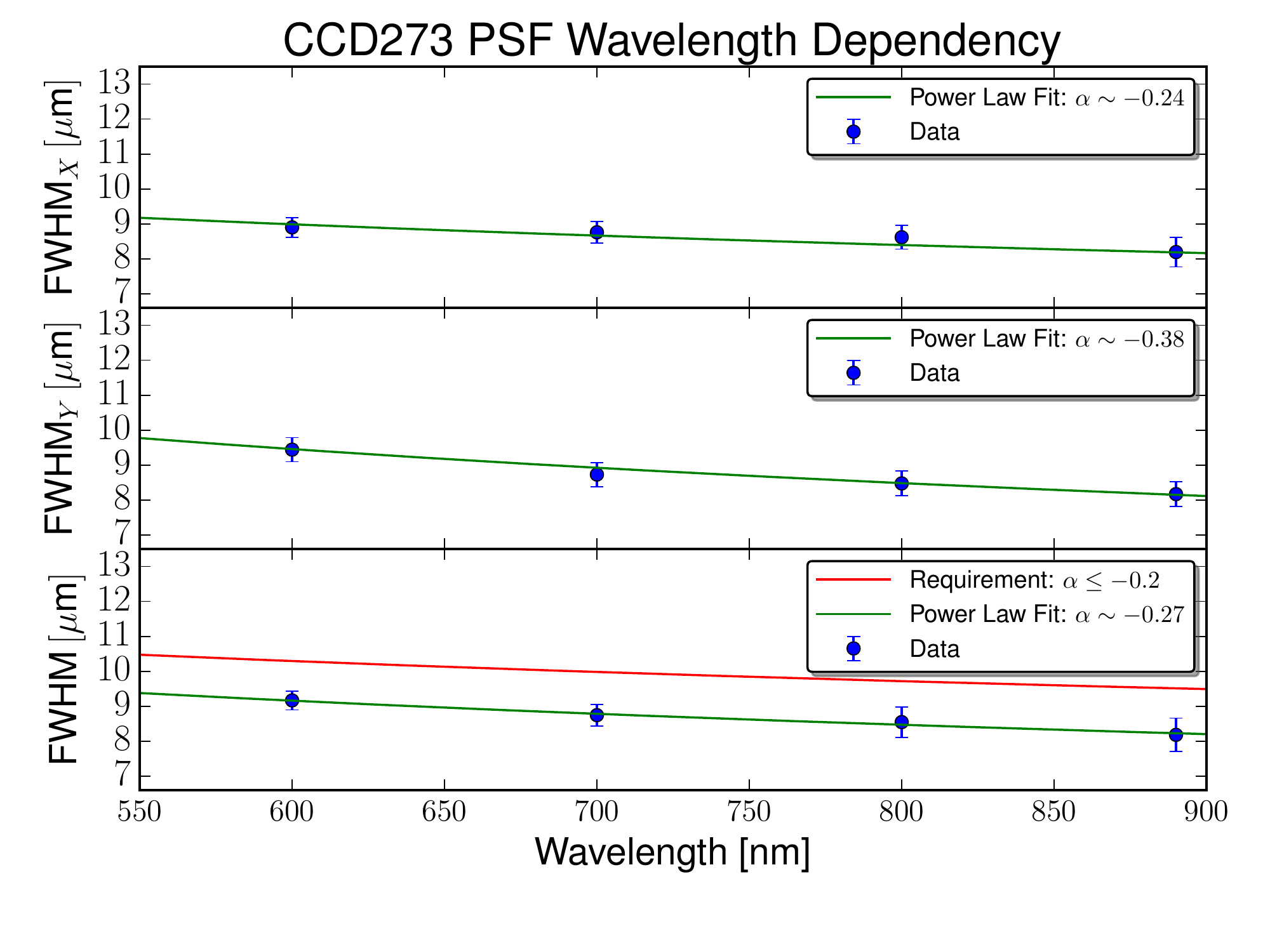}
\vspace{-10mm}
\caption{The CCD PSF size as a function of wavelength. For a back-illuminated CCD273 the CCD PSF narrows towards longer wavelengths. The trend can be approximated with a power law. The uncertainties shown are $3\sigma$ estimates from the MCMC chains.}
\label{fig:WavelengthDependency}  
\end{figure}

Figure \ref{fig:WavelengthDependency} shows the CCD273-84 PSF size as a function of wavelength. As expected, the PSF is larger at bluer wavelengths. However, the Figure shows that this change is not dramatic. If we parametrise the wavelength dependency with a power law: $\textrm{FWHM} \, \propto \, \lambda^{\alpha}$, we find that $\alpha \sim -0.27 \pm 0.03$ (the $1\sigma$ uncertainty has been derived from the MCMC chain). This is well within the requirement of $\textit{Euclid}$ VIS: $\alpha \leq -0.2$ (shown with red). The negative power, i.e. a narrower PSF towards redder wavelengths is important because it allows the detectors to counter the broadening of the optical PSF from the telescope, which grows as a function of wavelength. When the dependency of the detector PSF size as a function of wavelength is opposite to the dependency of the optical PSF, the system PSF size follows a weaker wavelength dependency. This is advantageous leading to a more uniform PSF size independent of the spectral energy density of the observed object.

The following equations describe the key characteristics of the intrinsic CCD273-84 PSF as a function of wavelength $\lambda$ (in nm):
\begin{eqnarray}
\mathrm{FWHM}_{x}(\lambda) & = & (41.0 \pm 9.1) \lambda^{(-0.237 \pm 0.05)} \\
\mathrm{FWHM}_{y}(\lambda) & = & (105.9 \pm 33.2) \lambda^{(-0.378 \pm 0.09)} \\
\mathrm{FWHM}(\lambda) & = & (52.2 \pm 7.8) \lambda^{(-0.272 \pm 0.03)} \\
e(\lambda) & = & (2.4 \pm 0.43) \lambda^{(-0.148 \pm 0.03)}\\
R^{2}(\lambda) & = & (55.8 \pm 21.7) \lambda^{(-0.513 \pm 0.06)} \quad .
\end{eqnarray}

\begin{figure}
\includegraphics[width=0.9\textwidth]{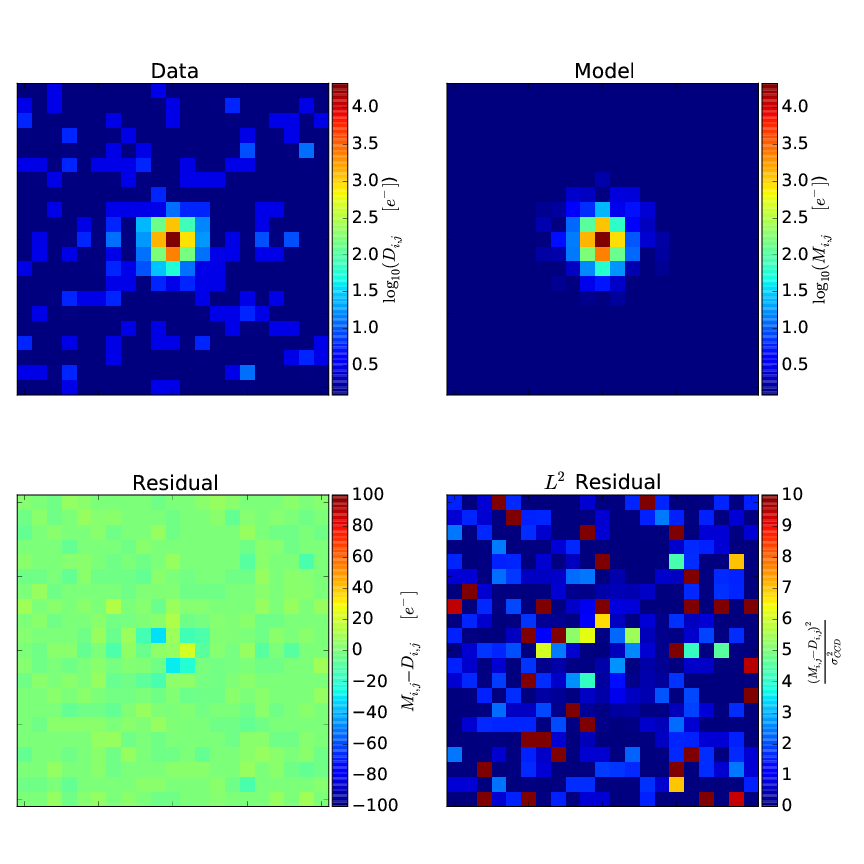}
\vspace{-3mm}
\caption{Comparison of the derived model $M_{i, j}$ against data $D_{i, j}$ taken at 600nm. An example of a non-joint fit shown is representative of the other fits at the same wavelength.}
\label{fig:600nmResiduals}  
\end{figure}

\begin{figure}
\includegraphics[width=0.9\textwidth]{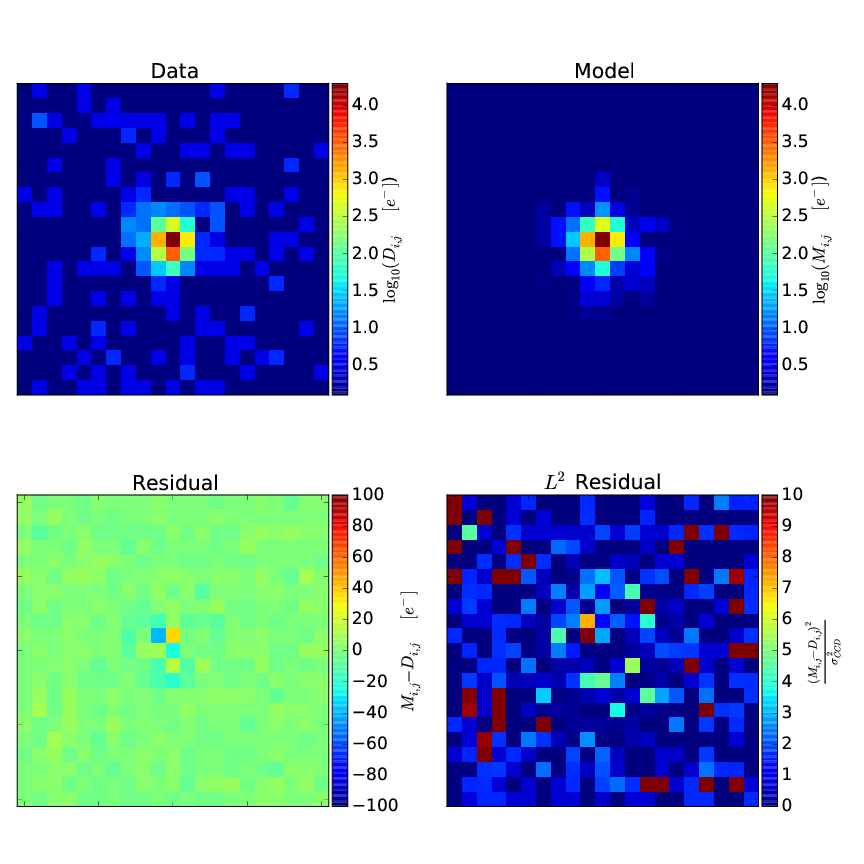}
\vspace{-3mm}
\caption{Comparison of the derived model $M_{i, j}$ against data $D_{i, j}$ taken at 700nm. An example of a non-joint fit shown is representative of the other fits at the same wavelength.}
\label{fig:700nmResiduals}  
\end{figure}

\begin{figure}
\includegraphics[width=0.9\textwidth]{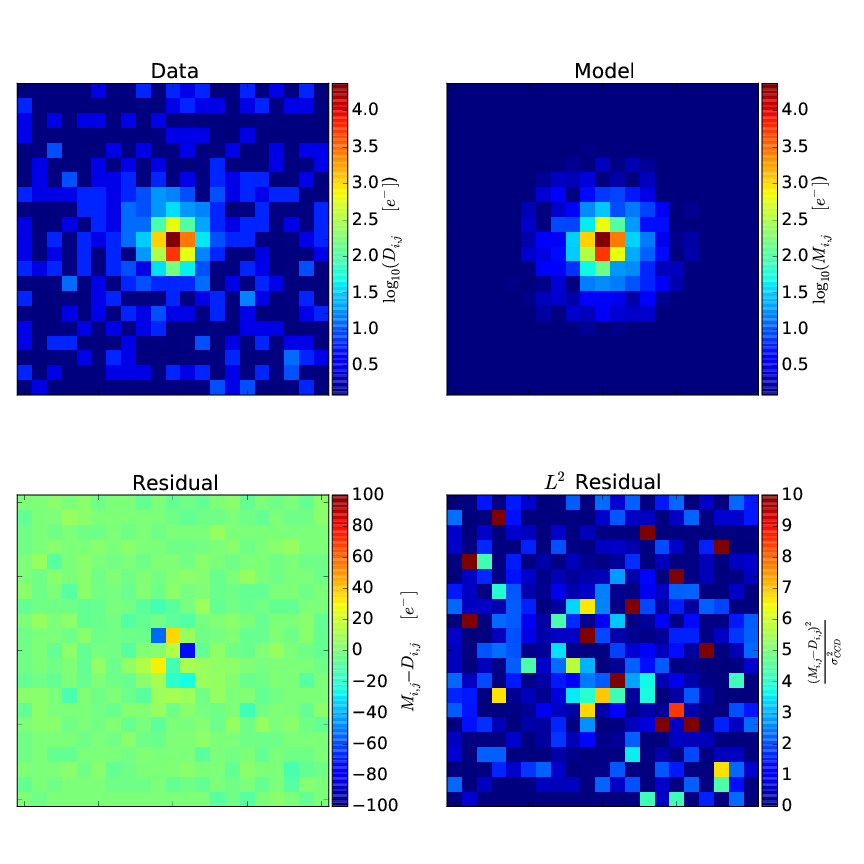}
\vspace{-3mm}
\caption{Comparison of the derived model $M_{i, j}$ against data $D_{i, j}$ taken at 800nm. An example of a non-joint fit shown is representative of the other fits at the same wavelength.}
\label{fig:800nmResiduals}  
\end{figure}

It should be noted that the PSF size wavelength dependency is steeper in the column (y/parallel) direction. This can be understood if the pixel barriers along the column direction set by electrodes generating an electric field are weaker than the barriers (separating the columns) set by doping. Indeed, Figure \ref{fig:WavelengthDependency} implies that these barriers are shallower than those set by doping in the serial (x) direction. This also means that the ellipticity of the CCD273-84 PSF is wavelength dependent: the PSF is slightly more elliptical at shorter wavelengths, reaching up to $\sim 7$ per cent by 550nm. However, this dependency is weaker than for the size measures, having a power $\sim -0.15$. Figures \ref{fig:600nmResiduals}, \ref{fig:700nmResiduals}, and \ref{fig:800nmResiduals}, show examples of the data, models, and residuals at 600, 700, and 800nm, respectively.

\subsection{Intensity Dependency}\label{ss:intensity}

Recently, several authors have reported results implying that CCD PSF sizes grow as a function of intensity \citep[e.g.][and references therein]{2014JInst...9C3048A, 2014JInst...9C4027R}. This effect has been dubbed as ``brighter-fatter'' and is often identified from pairwise flat field data showing that a photon transfer curve deviates from the shot noise prediction. However, it should be noted that these are indirect inferences about the CCD PSF from an even illumination (flatfield), and it is therefore important to derive the intensity dependency directly from spot-projected data.

\begin{figure}
\includegraphics[width=1.\textwidth]{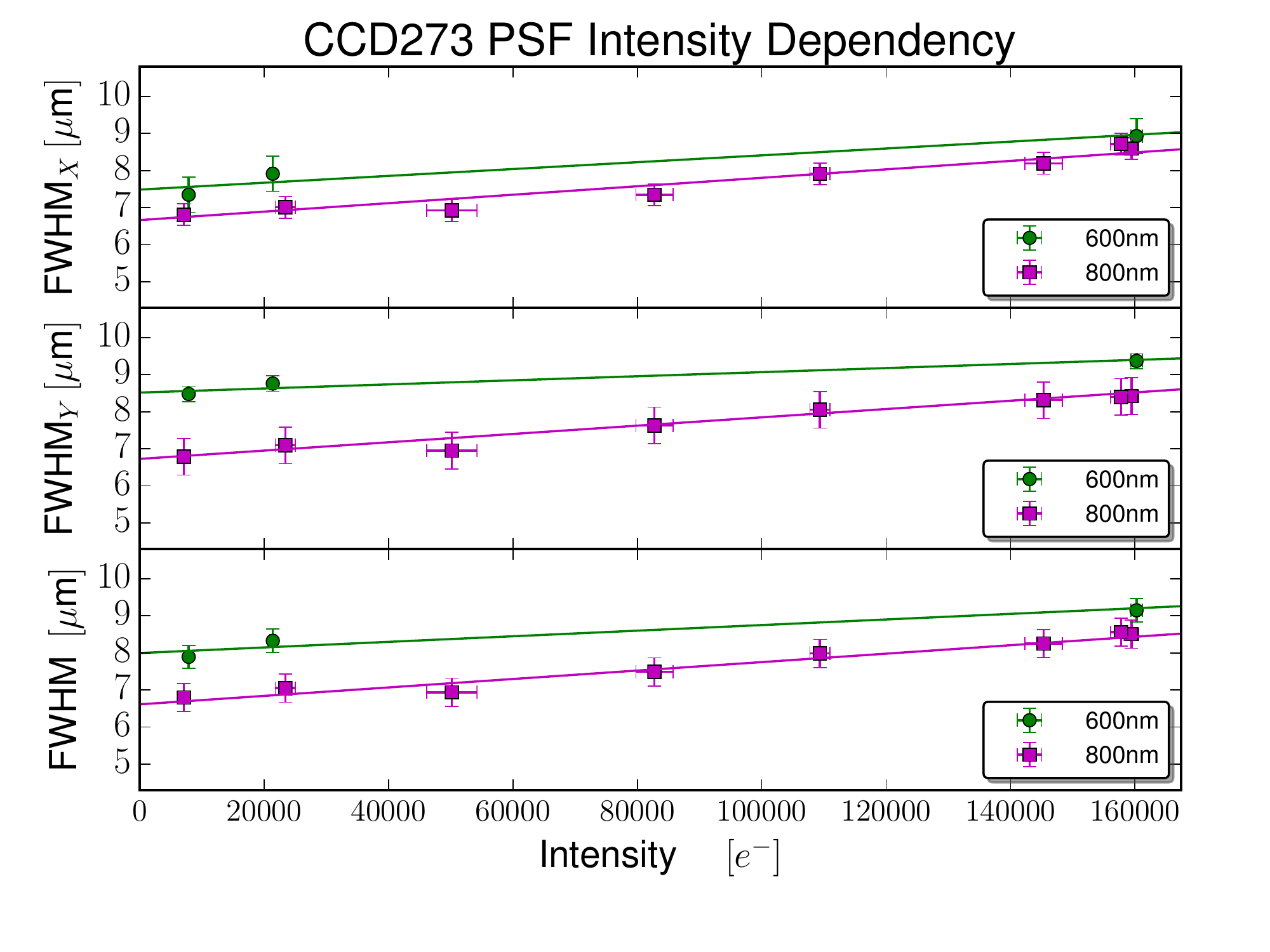}
\vspace{-10mm}
\caption{The CCD PSF size as a function of number of electrons in the peak pixel, dubbed as intensity, at two different wavelengths. The CCD PSF size grows as a function of intensity. The trend can be approximated with a linear correlation. The uncertainties shown are $3\sigma$.}
\label{fig:IntensityDependency}  
\end{figure}

Figure \ref{fig:IntensityDependency} shows the CCD273-84 PSF size as a function of intensity, defined here as the number of electrons in the peak pixel, at two different wavelengths centred at 600 and 800nm. The Figure shows that the PSF size grows as a function of intensity. This growth has a slightly different slope at 600nm from that at 800nm, especially in the column/parallel (y) direction. The slope of the CCD PSF FWHM at 600nm is $\sim 7.5\times 10^{-6}$ while it steepens to $\sim 1.1 \times 10^{-5}$ when measured at 800nm. This difference in the slope is potentially related to the fact that for a back-illuminated CCD the PSF size is larger at shorter wavelengths. If the pixel barriers are lower then it is possible that they are less influenced by the combined effect of the accumulated photoelectrons leading to a flatter intensity response. This explanation is consistent with the fact that we observe a flatter relation in the column/parallel direction (middle panel of Figure \ref{fig:IntensityDependency}). It should be noted however that at 600nm we measured the CCD PSF only at three different intensity levels and therefore the uncertainties in the correlation are higher.

At 800nm the intensity $I$ (in electrons) dependency can be described for the key characteristics as follows:
\begin{eqnarray}
\mathrm{FWHM}_{x}(I) & = & (6.66 \pm 0.21) +  (1.14 \pm 0.17)\times 10^{-5}I \\
\mathrm{FWHM}_{y}(I) & = & (6.72 \pm 0.12) +  (1.12 \pm 0.11)\times 10^{-5}I \\
\mathrm{FWHM}(I) & = & (6.61 \pm 0.09) +  (1.14 \pm 0.08)\times 10^{-5}I \\
e(I) & = & (8.54 \pm 4.0)\times 10^{-3} +  (10.26 \pm 3.83)\times 10^{-8}I\\
R^{2}(I) & = & (1.13 \pm 0.04) +  (4.16 \pm 0.27)\times 10^{-6}I 
\end{eqnarray}
It should be noted that when measured at 800nm the CCD PSF size dependency seems to have almost the same slope $(\sim 1.1 \times 10^{-5})$ in both serial and parallel direction. Thus, the intensity dependency of the CCD PSF ellipticity is very weak $(\sim 10^{-7})$, when measured at 800nm.

The signal dependent CCD PSF must be taken into account when building a system Point Spread Function model that incorporates both the optics and the detector effects. However, because the intensity dependency is linear  with intensity (see Fig. \ref{fig:IntensityDependency}) this can be taking into account by introducing a term to the system PSF model that broadens the detector PSF with increasing intensity.

\subsection{CCD273 PSF Model}

Results of the previous sections have shown that a two-dimensional Gaussian can describe the intrinsic CCD273 PSF, which depends both on wavelength and intensity of the illumination. Taking the results of Sections \ref{ss:wavelength} and \ref{ss:intensity} we can now write mathematical descriptions for the key properties as a function of these as follows:
\begin{eqnarray}
\mathrm{FWHM}_{x}(I, \lambda) & = & \frac{\mathrm{FWHM}_{x}(I)}{\mathrm{FWHM}_{x}(I_{0})}\mathrm{FWHM}_{x}(\lambda) \\
\mathrm{FWHM}_{y}(I, \lambda) & = & \frac{\mathrm{FWHM}_{y}(I)}{\mathrm{FWHM}_{y}(I_{0})}\mathrm{FWHM}_{y}(\lambda) \\
\mathrm{FWHM}(I, \lambda) & = & \frac{\mathrm{FWHM}(I)}{\mathrm{FWHM}(I_{0})}\mathrm{FWHM}(\lambda) \\
e(I, \lambda) & = & \frac{e(I)}{\mathrm{e(I_{0})}}e(\lambda) \\
R^{2}(I, \lambda) & = & \frac{R^{2}(I)}{\mathrm{R^{2}(I_{0})}}R^{2}(\lambda) \quad ,
\end{eqnarray}
where the intensity zero point $I_{0} \sim 157800$. It should be noted that here we have assumed that the intensity relation is independent of the wavelength. 

\section{Summary and Conclusions} \label{sec:summary}

We have developed a novel modelling method to measure the intrinsic Charge-Coupled Device (CCD) Point Spread Function (PSF) characteristics. In our method the optical spot projection system and the CCD PSF were modelled jointly. Bayesian posterior probability density function sampling was used to fit the model to available data. In this paper the method is applied to spot projected data collected with a back-illuminated development \Euclid Visible Instrument CCD273-84.

Using simulated data the generative model fitting was shown to allow good parameter estimations even when these data are not well sampled. After confirming the parameter recovery the intrinsic CCD273-84 PSF was characterised as a function of wavelength and intensity. The CCD PSF kernel size was found to increase with increasing intensity and decreasing wavelength. The pre-development CCD273 tested was found to meet all the \Euclid requirements with comfortable margins. Finally, mathematical descriptions for the CCD PSF size, both Full Width at Half Maximum and $R^{2}$, and ellipticity were provided to enable the modelling of the CCD PSF.

\begin{acknowledgements}
SMN would like to thank the \Euclid CCD Working Group members, Gary Bernstein, and Robert Lupton for useful comments and stimulating discussions. We thank the anonymous reviewer for comments that helped us to improve the paper. The authors acknowledge the Euclid Consortium, the European Space Agency and the support of a number of agencies and institutes that have supported the development of Euclid. A detailed complete list is available on the Euclid web site (\url{http://www.euclid-ec.org/}). In particular the Agenzia Spaziale Italiana, the Centre National d’Etudes Spatiales, the Deutches Zentrum fur Luft- and Raumfahrt, the Danish Space Research Institute, the Fundação para a Ciênca e a Tecnologia, the Ministerio de Economia y Competitividad, the National Aeronautics and Space Administration, the Netherlandse Onderzoekschool Voor Astronomie, the Norvegian Space Center, the Romanian Space Agency, the United Kingdom Space Agency and the University of Helsinki.
\end{acknowledgements}

\bibliographystyle{aps-nameyear}      
\bibliography{example}                
\nocite{*}

\end{document}